# Topological Kagome magnet $Co_3Sn_2S_2$ thin flakes with high electron mobility and large anomalous Hall effect

M. Tanaka[1*†], Y. Fujishiro[1*†], M. Mogi[1], Y. Kaneko[2], T. Yokosawa[1],
N. Kanazawa[1], S. Minami[3], T. Koretsune[4], R. Arita[1,2],
S. Tarucha[2], M. Yamamoto[2] and Y. Tokura[1,2,5]

[1] Department of Applied Physics, University of Tokyo, Bunkyo-ku, Tokyo 113-8656, Japan

[2] RIKEN Center for Emergent Matter Science (CEMS), Wako, Saitama 351-0198, Japan

[3] Department of Physics, University of Tokyo, Hongo, Tokyo 113-0033, Japan

[4] Department of Physics, Tohoku University, Aoba-ku, Sendai 980-8578, Japan

[5] Tokyo College, University of Tokyo, Bunkyo-ku, Tokyo 113-8656, Japan

† These authors equally contributed to this work.

* To whom correspondence should be addressed.

E-mail: miuko.t@gmail.com and fujishiro@cmr.t.u-tokyo.ac.jp

**Magnetic Weyl semimetals attract considerable interest not only for their topological quantum phenomena but also as an emerging materials class for realizing quantum anomalous Hall effect in the two-dimensional limit. A shandite compound $Co_3Sn_2S_2$ with layered Kagome-lattices is one such material, where vigorous efforts have been devoted to synthesize the two-dimensional crystal. Here we report a synthesis of $Co_3Sn_2S_2$ thin flakes with a thickness of 250 nm by chemical vapor transport method. We find that this facile bottom-up approach allows the formation of large-sized $Co_3Sn_2S_2$ thin flakes of high-quality, where we identify the largest electron mobility (~2,600 $cm^2V^{-1}s^{-1}$) among magnetic topological semimetals, as well as the large anomalous Hall conductivity (~1,400 $\Omega^{-1}cm^{-1}$) and anomalous Hall angle (~32 %) arising from the Berry curvature. Our study provides a viable platform for studying high-quality thin flakes of magnetic Weyl semimetal and stimulate further research on unexplored topological phenomena in the two-dimensional limit.**

Magnetic Weyl semimetals (WSMs) with broken time-reversal symmetry exhibit unique physical properties arising from the interplay between magnetism and band topology [1-5]. In addition to the large magnetoresistance and high carrier mobility often observed in topological semimetals [6-8], the large anomalous Hall conductivity and anomalous Hall angle at zero magnetic field are known to be the hallmarks of diverging Berry curvature at the Weyl points in the presence of intrinsic magnetism [1,9-12]. One other emerging aspect of the magnetic WSMs is that they are a new class of materials that can potentially realize the quantum anomalous Hall effect (QAHE) [13,14] in the two-dimensional (2D) limit. Since the Weyl fermions are only defined in the three-dimensional (3D) momentum space, the breaking of the translational symmetry along one direction can induce a topologically non-trivial gap by quantum confinement effect [11]. The QAHE shows up when the Fermi level ($E_F$) is tuned inside the gap. Furthermore, if the intrinsic magnetism realizes a large gap with a high magnetic ordering temperature, the QAHE may occur at a high temperature, offering versatile applications ranging from dissipationless electronics to topological quantum computing [15,16].

A ferromagnetic shandite compound $Co_3Sn_2S_2$ has attracted particular interest to this end. $Co_3Sn_2S_2$ crystallizes in a rhombohedral space group *R-3m* (No. 166) [17], where magnetic Co atoms form a 2D Kagome lattice with an Sn atom at the center of the hexagon, which is sandwiched between two S layers (Fig. 1(a)). These [S-($Co_3Sn$)-S] structures are further connected by the hexagonal Sn layers. In this compound, Weyl points exist close to the $E_F$ with almost no interference from trivial bands, leading to the emergence of large anomalous Hall effect and the anomalous Hall angle [18-20]. In addition to the quasi-2D layered crystal structure as well as the simple semi-metallic band structure, the intrinsic magnetism with the high Curie temperature ($T_c \sim 180$ K) [21,22] makes it an ideal platform for exploring high-temperature QAHE [23].

Recently, there has been much effort devoted to the synthesis of 2D crystals of $Co_3Sn_2S_2$ from various approaches. In addition to the top-down method of microstructure fabrication using a focused ion beam (thickness ~350 nm) [24], the bottom-up approach such as the thin-film growth by molecular beam epitaxy (MBE) (thickness ~ 18 nm) [25] or sputtering deposition (thickness ~35 nm) [26] has been reported. However, the quality or uniformity of the thin film is limited due to the formation of nano-sized grains, especially when the thickness becomes small [25]. While the QAHE expected only in the 2D limit (less than three unit-cells), it would be important to establish various feasible approaches to synthesize 2D crystals of $Co_3Sn_2S_2$ of high-quality, which may allow us to study the non-trivial crossover behaviors of topological transport properties from the bulk state towards the 2D limit.

Here we report an alternative bottom-up method for growing thin crystals of $Co_3Sn_2S_2$ using chemical vapor transport (CVT). While the CVT method is commonly used as bulk crystal growth approach, it can also be extended to the fabrication of single-crystalline nanoflakes, as demonstrated in several layered materials including transition metal dichalcogenides [27-29]. Compared to the thin film growth by vapor deposition of each element, the CVT method may potentially realize more high-crystalline and uniform sample by using the target material as a vapor source. In this work, we report the successful synthesis of high-quality $Co_3Sn_2S_2$ thin flakes with a thickness of 250 nm, exhibiting high electron mobility (~2,600 $cm^2V^{-1}s^{-1}$) as well as large anomalous Hall effect (AHE) comparable or larger than those in bulk samples. The high electron mobility leads to an unusual shape of the anomalous Hall hysteresis with "dispersive-resonance" profile [8], which has been rarely reported in magnetic materials. The emergence of the high electron mobility and the large AHE is discussed in terms of the possible hole-doping in the thin flake, which is corroborated by variation of carrier density estimated from the band structure calculations. Our study provides a simple and facile approach for studying topological transport properties in high-quality $Co_3Sn_2S_2$ thin flakes, which may facilitate further research on the realization of QAHE in the 2D limit.

We grew $Co_3Sn_2S_2$ thin flakes with the CVT method as schematically illustrated in Fig. 1(b). The source material is the milled powder of single-crystalline $Co_3Sn_2S_2$ grown by the Bridgman method, which was sealed in an evacuated quartz tube with iodine added as a transport agent. Then, the growth of the thin flakes was performed in a temperature gradient from 950 °C to 850 °C for 120 hours. The thin flakes formed at the inner surface of the quartz tube were picked up by the thermal release tape and subsequently released onto the SiO$_2$/Si substrate upon heating at 120 °C (Fig. 1(c)). The as-grown thin flakes with a thickness of 50 - 250 nm were formed at a high density as observed by optical microscopy (Figs. 1(d)-(f)), where the thickness was identified by atomic force microscopy (AFM). It is noteworthy that even though Co$_3$Sn$_2$S$_2$ has a relatively strong interlayer coupling unlike the van-der-Waals 2D crystals, the obtained flakes have a large area with respect to the thickness (*e.g.*, few hundreds of micro-meters size for a thickness of 50-250 nm).

The gold electrodes for the transport measurement were successfully fabricated onto the 250 nm-thick flakes as shown in Fig. 2(a) (thinner flakes did not have large enough size for the device fabrication at this moment) by electron beam (EB) lithography with a wet etching approach (see Methods for details). In our device fabrication process, thin flakes are never exposed to the resist, which we found to be important in making electrical contacts. Figure 2(b) shows the temperature (*T*) dependence of longitudinal resistivity ($\rho_{xx}$) of the thin flake (blue line) and the bulk single crystal grown by the Bridgman method (black line), both of which exhibit the kink structure at the $T_c \sim 177$ K. On the other hand, the observed residual resistivity ($\rho_{2K} \sim 23$ μΩcm) for the thin flake is smaller than that of the bulk sample ($\rho_{2K} \sim 91$ μΩcm) or those in previous studies [18,19,30], suggesting that the thin flake has high sample quality with few disorders. Accordingly, a large residual resistivity ratio (RRR = $\rho_{300K}/\rho_{2K}$) of 20 is also identified for the thin flake. Here, $\rho_{300K}$ and $\rho_{2K}$ are the $\rho_{xx}$ at $T = 300$ K and 2 K, respectively. We note that this high sample quality partly contributes to the emergence of high electron mobility (~2,600 $cm^2V^{-1}s^{-1}$) in the thin flake which will be discussed later, in a sense that disorders set an upper limit of the mobility [30]. We also observe large non-saturating magnetoresistance (MR) reaching ~180 % at perpendicular magnetic field of $B = 14$ T at $T = 2$ K in the thin flake (Fig. 2(c)). Here, MR is defined as $[(\rho_{xx}(B) - \rho_{xx}(0)]/\rho_{xx}(0)$ where $\rho_{xx}(B)$ is the longitudinal resistivity in an applied magnetic field $B$. The observed large MR in the thin flake cannot be fully explained by the electron-hole compensation mechanism [6,7] and requires further investigations (see Supplementary Fig. 1 for details). The Hall resistivity shows a field-nonlinear behavior typical of multi-carrier systems, where the distinct difference between the thin flake and the bulk sample suggests the variation in electron/hole carrier density (Fig. 2(c)). In addition, the thin flake shows a large coercive field ($H_c$) which is enhanced by a factor of 18 at $T = 2$ K ($\mu_0H_c = 5.5$ T) compared to that in bulk samples ($\mu_0H_c = 0.3$ T) (Fig. 2(d)). The similar enhancement of the coercivity has been also reported for $Co_3Sn_2S_2$ thin films [25,26].

Figure 3(a) shows the $B$-dependence of the Hall conductivity ($\sigma_{xy}$) measured at various temperatures for the 250 nm-thick flake. At $T = 140$ K, $\sigma_{xy}$ exhibits a rectangular hysteresis loop with a sharp magnetization reversal associated with the AHE. While the AHE (the yellow shaded region in Fig. 3(a)) is almost constant for the variation of $B$ at $T = 140$ K, a field-nonlinear behavior becomes pronounced upon lowering the temperature, which we assign to the ordinary Hall effect (OHE). Such a notable non-linear behavior is absent in the bulk sample (Supplementary Fig. 2). To obtain the mobility and carrier density from the OHE, we fitted the $\sigma_{xy}$ with a two-carrier Drude model (see Methods), after subtracting the field-independent $\sigma^A_{xy}$ [=$\sigma_{xy}(B=0$ T)] value at every magnetic field point. As shown by the dashed line at $T = 2$ K in Fig. 2(a), the Drude fitting can well reproduce the nonlinear magnetic field profile of $\sigma_{xy}$ with the parameter values: $\mu_e = 2579$ cm$^2$V$^{-1}$s$^{-1}$, $\mu_h = 159$ cm$^2$V$^{-1}$s$^{-1}$, $n_e = 1.53 \times 10^{19}$ cm$^{-3}$, and $n_h = 3.37 \times 10^{20}$ cm$^{-3}$ (Table 1). Here, $\mu_e$ and $\mu_h$ are the mobility while $n_e$ and $n_h$ are the carrier density of electron and hole, respectively. The observed "dispersive-resonance" profile with a sharp peak in the OHE can only be seen in high-mobility systems [8]. We note that such a coexistence of the AHE hysteresis loop and the dispersive-resonance profile of the OHE has been rarely observed; indeed, the electron mobility observed here is the largest value ever reported for topological semimetals with intrinsic magnetism (*e.g.*, GdPtBi (~1,500 cm$^2$V$^{-1}$s$^{-1}$) [31], MnBi$_2$Te$_4$ flake (~1,100 cm$^2$V$^{-1}$s$^{-1}$) [32], Co$_2$MnGa (~35 cm$^2$V$^{-1}$s$^{-1}$) [33]), suggesting a presence of small electron pockets in addition to the high sample quality of the thin flake.

The $T$ dependence of the $\sigma^A_{xy}$ taken at $B = 0$ T and the anomalous Hall angle defined as $\sigma^A_{xy}/\sigma_{xx}$ are shown in Fig. 3(b), where $\sigma_{xx}$ is the longitudinal conductivity. The $\sigma^A_{xy}$ reaching ~1,400 $\Omega^{-1}$cm$^{-1}$ is nearly independent of temperature below $T = 160$ K, which suggests the Berry-curvature mechanism. The observed $\sigma^A_{xy}$ is comparable or slightly enhanced compared to those in previous studies, which reported the $\sigma^A_{xy}$ of 500－1,400 $\Omega^{-1}$cm$^{-1}$ [18,19,24,25,30]. Accordingly, the large anomalous Hall angle reaching ~ 32 % around $T = 140$ K was identified in the thin flake. To investigate the possible extrinsic contributions to the AHE, we adopted the established scaling relation between $\rho_{xx}$ and $\rho_{yx}$ for the AHE [10]. The Hall resistivity at zero magnetic field can be written as $\rho_{yx} = (\alpha \rho_{xx} + \beta \rho_{xx}^2) \cdot M$, where the first term represents the skew-scattering while the second term corresponds to the sum of side-jump and intrinsic contributions. Both parts are linearly proportional to magnetization ($M$). Hence, the linear fitting for $\rho_{yx}/\rho_{xx}M$ against $\rho_{xx}$ gives the parameters $\alpha$ (intercept) and $\beta$ (slope). As shown in Fig. 3(c), the linear fitting reveals a negligible contribution from the skew-scattering, being consistent with the previous reports in bulk samples [19,34]. Further analysis also confirms the small contribution from the side-jump being less than 100 $\Omega^{-1}cm^{-1}$ (see Supplementary Fig. 3), hence the observed large AHE is dominated by the Berry-curvature from the momentum space.

To understand the emergence of high electron mobility as well as the large intrinsic AHE in a comprehensive way, we performed the band structure calculations of $Co_3Sn_2S_2$ (Fig. 4(a)). The electronic band structure of $Co_3Sn_2S_2$ without the spin-orbit coupling (SOC) has two sets of linear band crossing points of the nodal ring along the Γ－L and L－U paths, which are located slightly above and below the $E_F$, respectively. The Weyl points appear at 60 meV above the $E_F$ by including the SOC, while the nodal ring forms a gap due to anti-crossings (the inset of Fig. 4(a)), resulting in the formation of multiple Fermi surfaces [20]. In undoped samples, the carrier density as well as the mobility for electron and hole become roughly equal as reported in bulk samples (Table 1) [18,30]. On the other hand, the reduction(enhancement) of electron(hole) carrier density or the enhancement(reduction) of electron(hole) mobility observed in the thin flake (Table 1) can be attributed to the effective process of hole-doping (Fig. 4(b), see also Supplementary Fig. 4 for Fermi pockets at various energy). Indeed, the observed electron(hole) carrier density is consistent with the calculated values for the $E_F$ shift by about -30 meV(-20 meV) or the change in electron(hole) density by ~$1.4\times10^{20}$ cm$^{-3}$ (~$1.6\times10^{20}$ cm$^{-3}$) (Fig. 4(b)). The reason for the possible hole-doping remains unclear at this moment, and further exploration of the synthesis conditions for CVT and device fabrication process would be necessary to reveal its origin. At the same time, we find that the $E_F$ shift to hole-doping side by 20 meV or the increase of the hole density by ~$1.6\times10^{20}$ cm$^{-3}$ corresponds to the maximum region for the calculated $\sigma^A_{xy}$ (Fig. 4(c)), which corroborates our observation of the large AHE being comparable or slightly enhanced than in undoped bulk samples. A similar enhancement of the $\sigma^A_{xy}$ is also confirmed in In-substituted (*i.e.,* hole-doped) bulk samples, which is attributed to the distribution of gapped nodal ring structure which gives the hot zone of the Berry curvature below the Fermi level [34]. In the meanwhile, the reason why the observed $\sigma^A_{xy}$ in the thin flake is even larger than that of the calculated value (1,150 $\Omega^{-1}$cm$^{-1}$) remains elusive. Although the thickness of 250 nm belongs to the bulk regime, we speculate that there may be already some non-trivial affects from the reduced dimensionality, such as the modification of band structures or contributions from the surface states. Systematic investigations in thinner flakes would be imperative to reveal such a possibility.

In summary, we have established an accessible arena for studying high-quality $Co_3Sn_2S_2$ thin flakes by using the CVT method. The quasi-two-dimensional crystal structure of $Co_3Sn_2S_2$ has successfully led to the formation of large-sized thin flakes, despite the presence of relatively strong interlayer coupling. Towards the synthesis of even thinner flakes, it would be important to optimize the growth parameters of CVT [26-28]. In particular, the growth rate needs to be suppressed by varying transport agent and temperature gradient while the growth time should be shortened. The CVT-grown thin flake exhibits the highest mobility among topological semimetals with intrinsic magnetism, as well as the large intrinsic AHE arising from the Berry curvature. These results suggest the high sample quality and uniformity of the thin flake, which would be important for studying exotic quantum phenomena of the magnetic Weyl semimetal in reduced dimensionality, and paves the way for the realization of high-temperature QAHE in the 2D limit.

**References**

[1] Murakami, S. Phase transition between the quantum spin Hall and insulator phases in 3D: emergence of a topological gapless phase. *New J. Phys.* **9**, 356 (2007).

[2] Wan, X. Turner, A. M., Vishwanath, A. & Savrasov, S. Y. Topological semimetal and Fermi-arc surface states in the electronic structure of pyrochlore iridates. *Phys. Rev. B* **83**, 205101 (2011).

[3] Burkov, A. A. & Balents, L. Weyl semimetal in a topological insulator multilayer. *Phys. Rev. Lett.* **107**, 127205 (2011)

[4] Armitage, N. P., Mele, E. J. & Vishwanath, A. Weyl and Dirac semimetals in three-dimensional solids. *Rev. Mod. Phys.* **90**, 015001 (2018)

[5] Nagaosa, N., Morimoto, T. & Tokura, Y. Transport, magnetic and optical properties of Weyl materials. *Nat. Rev. Mater.* Advance online publication. (2020) <doi:10.1038/s41578-020-0208-y>

[6] Ali, M. N. *et al*. Large non-saturating magnetoresistance in $WTe_2$. *Nature* **514**, 205 (2014)

[7] Shekhar, C. *et al*. Extremely large magnetoresistance and ultrahigh mobility in the topological Weyl semimetal candidate NbP, *Nat. Phys.* **11**, 645 (2015)

[8] Liang, T. *et al*. Ultrahigh mobility and giant magnetoresistance in the Dirac semimetal $Cd_3As_2$, *Nat. Phys.* **11**, 645 (2015)

[9] Fang, Z. *et al.* The anomalous Hall effect and magnetic monopoles in momentum space. *Science* **302**, 92–95 (2003)

[10] Nagaosa, N., Sinova, J., Onoda, S., MacDonald, A. H. & Ong, N. P. Anomalous Hall effect, *Rev. Mod. Phys.* **82**, 1539 (2010)

[11] Xu, G., Weng, H., Wang, Z., Dai, X. & Fang, Z. Chern semimetal and the quantized anomalous Hall effect in HgCr$_2$Se$_4$, *Phys. Rev. Lett.* **107**, 186806 (2011)

[12] Burkov, A. A. Anomalous Hall effect in Weyl metals. *Phys. Rev. Lett.* **113**, 187202 (2014).

[13] Yu, R. e*t al*. Quantum anomalous Hall effect in magnetic topological insulators. *Science* **329**, 5987 (2010)

[14] Chang, C. -Z. e*t al*. Experimental observation of the quantum anomalous Hall effect in a magnetic topological insulator. *Science* **340**, 167 (2013)

[15] Halperin, B. I. Quantized Hall conductance, current-carrying edge states, and the existence of extended states in a two-dimensional disordered potential. *Phys. Rev. B* **25**, 2185 (1982)

[16] Qi, X. -L., Hughes, T. L. & Zhang, S. -C. Chiral topological superconductor from the quantum Hall state. *Phys. Rev. B* **82**, 184516 (2010)

[17] Weihrich, R. & Anusca, I & Zabel, M. Half-antiperovskites: Structure and type-antitype relations of Shandites M$_{3/2}$AS (M = Co, Ni; A = In, Sn). *Z. Anorg. Allg. Chem.* **631**, 1463 (2005)

[18] Liu, E. *et al*. Giant anomalous Hall effect in a ferromagnetic kagome-lattice semimetal. *Nat. Phys.* **14**, 1125 (2018)

[19] Wang, Q. e*t al*. Large intrinsic anomalous Hall effect in half-metallic ferromagnet Co$_3$Sn$_2$S$_2$ with magnetic Weyl fermions. *Nat. Commun.* **9**, 3681 (2018)

[20] Liu, D. F. *et al*. Magnetic Weyl semimetal phase in a Kagomé crystal. *Science* **365**, 1282 (2019)

[21] Vaqueiro, P. & Sobany, G. G. A powder neutron diffraction study of the metallic ferromagnet $Co_3Sn_2S_2$. *Sol. Stat. Sci.* **11**, 513 (2009)

[22] Schnelle, W. *et al*. Ferromagnetic ordering and half-metallic state of $Sn_2Co_3S_2$ with the Shandite-type structure. *Phys. Rev. B* **88**, 144404 (2013)

[23] Muechler, L. *et al*. Emerging chiral edge states from confinement of a magnetic Weyl semimetal in $Co_3Sn_2S_2$. *Phys. Rev. B* **101**, 115106 (2020)

[24] Geishendorf, K. *et al*. Magnetoresistance and anomalous Hall effect in micro-ribbons of the magnetic Weyl semimetal $Co_3Sn_2S_2$. *Appl. Phys. Lett.* **114**, 092403 (2019)

[25] Li, S. *et al*. Epitaxial growth and transport properties of magnetic Weyl semimetal $Co_3Sn_2S_2$ thin films. *ACS Appl. Electron. Mater.* **2**, 126 (2020)

[26] Fujiwara, K. *et al*. Ferromagnetic $Co_3Sn_2S_2$ thin films fabricated by co-sputtering. *Jpn. J. Appl. Phys.* **58**, 050912 (2019)

[27] Wang, J. *et al*. Controlled synthesis of two-dimensional 1T-$TiSe_2$ with charge density wave transition by chemical vapor transport. *J. Am. Chem. Soc.* **138**, 16216 (2016)

[28] Hu, D. *et al*. Two-dimensional semiconductor grown by chemical vapor transport. *Angew. Chem. Int. Ed.* **56**, 1 (2017)

[29] Grönke, M. *et al*. Chemical vapor growth and delamination of α-$RuCl_3$ nanosheets down to the monolayer limit. *Nanoscale.* **10**, 19014 (2018)

[30] Ding, L. *et al*. Intrinsic anomalous Nernst effect amplified by disorder in a half-metallic semimetal. *Phys. Rev. X.* **9**, 041061 (2019)

[31] Hirschberger, M. *et al*. The chiral anomaly and the thermopower of Weyl fermions in the half-Heusler GdPtBi. *Nat. Matter.* **15**, 1161 (2015)

[32] Lee, S. H. *et al*. Transport evidence for a magnetic-field induced ideal Weyl state in antiferromagnetic topological insulator $Mn(Bi_{1-x}Sb_x)_2Te_4$. arXiv:2002.10683

[33] Belopolski, I. *et al*. Discovery of topological Weyl fermion lines and drumhead surface states in a room temperature magnet. *Science* **365**, 1278 (2019)

[34] Zhou, H. *et al*. Enhanced anomalous Hall effect in the magnetic topological semimetal $Co_3Sn_{2-x}In_xS_2$. *Phys. Rev. B* **101**, 125121 (2020)

[35] Perdew, J., Burke, K. & Ernzerhof, M. Generalized gradient approximation made simple. *Phys. Rev. Lett.* **77**, 3865 (1996)

[36] Giannozzi, P. *et al*. Quantum espresso: a modular and open-source software project for quantum simulations of materials. *J. Phys. Condens. Matter* **21**, 395502 (2009)

[37] Blöchl, P. E. Projector augmented-wave method. *Phys. Rev. B* **50**, 17953 (1994)

[38] Pizzi, G. *et al*. Wannier90 as a community code: new features and applications. *J. Phys. Condens. Matter* **32**, 165902 (2020)

[39] Wu, Q. *et al*. WannierTools : An open-source software package for novel topological materials, *Comp. Phys. Comm.* **224**, 405 (2018)

**METHODS**

**CVT growth of thin flakes**

We used the milled powder of the single crystalline $Co_3Sn_2S_2$ (~300 mg), which was grown by the Bridgman method, as a starting material for CVT. The iodine (~10 mg) was added as a transport agent and then sealed in an evacuated quartz tube at a pressure of

<1×10$^{-4}$ Pa. The sealed quartz tube with a diameter of 9 mm and a length of ~13 cm was put in the three-zone furnace. The CVT growth was performed in a temperature gradient from 950 °C to 850 °C for 120 hours, after the pretreatment with the inverted temperature gradient for 12 hours.

**Device fabrication**

Thin flakes were transferred from the quartz tube onto 285 nm SiO$_2$/Si substrates by using a thermal release tape, followed by the direct deposition of 600 nm-thick gold by electron beam evaporators. Then, the electrodes in a Hall bar geometry were patterned onto the thin flake through electron beam lithography with poly (methyl methacrylate) (PMMA) as an etching mask for the subsequent wet-etching process. The etching of gold was performed by immersing in undiluted potassium iodine for 44 seconds. After cleaning the sample with running purified water for ~7 minutes, the PMMS resist was removed by immersing in acetone.

**Transport measurement**

Magneto-transport measurements were performed in a Quantum Design PPMS with a standard four-probe method. The magnetic field was applied along the *c*-axis of the sample and perpendicular to the electric current. The longitudinal conductivity ($\sigma_{xx}$) and the Hall conductivity ($\sigma_{xy}$) were calculated as $\sigma_{xx} = \rho_{xx}/(\rho_{xx}^2 + \rho_{yx}^2)$ and $\sigma_{xy} = \rho_{yx}/(\rho_{xx}^2 + \rho_{yx}^2)$. Here, $\rho_{xx}$ and $\rho_{yx}$ are the longitudinal and Hall resistivity, respectively. The fitting by the two-carrier Drude model was applied to obtain mobilities and carrier densities at low temperatures : $\sigma_{xy} = \frac{\mu_e^2 n_e eB}{1+\mu_e^2 B^2} + \frac{\mu_h^2 n_h eB}{1+\mu_h^2 B^2}$. Here, $\mu_e$ and $\mu_h$ represent mobility of electron and hole, respectively, while $n_e$ and $n_h$ represent the carrier density of electron and hole, respectively.

**Band structure calculation**

Electronic structure calculations were performed within the generalized gradient approximation [35] in the framework of the density functional theory as implemented in the quantum-ESPRESSO package [36]. The plane wave basis sets with projector augmented wave (PAW) scheme [37] was used. For the Berry curvature and the anomalous Hall conductivity calculations, a Wannier-interpolated band structure was employed [38,39]. The peak position of the calculated $\sigma_{xy}^A$ as a function of $E_F$ varies between experimental and optimized lattice parameters (see Supplementary Fig. 5 for a detailed discussion).

## AUTHOR CONTRIBUTIONS

M.T., Y.F., and M.M. conceived the experiment. Y.F. grew thin flakes with Y.K.. M.T. fabricated the electrodes with support from T.Y.. Y.F. performed transport measurements and analyzed the data with the support from M.T., M.M., and N.K.. S.M., T.K., and R.A. performed the band structure calculations. M.T. and Y.F. wrote the manuscript and all the authors discussed the results. M.Y., S.T., and Y.T. supervised the project.

## ACKNOWLEDGMENT

We thank T. Liang, A. Kikkawa, Y. Nakagawa, and M. Onga for experimental supports and R. Yamada, Y. Okamura, M. Hirschberger, R. Yoshimi, A. Tsukazaki, and M. Kawasaki for fruitful discussions. M. T. and Y. F. are supported by the Materials Education program for the future leaders in Research, Industry, and Technology (MERIT). The work was supported in part by JST CREST (Grants No. JPMJCR16F1 and No. JPMJCR1874) and JSPS KAKENHI (Grants No. 18J20959 and Grants No. 19H00650).

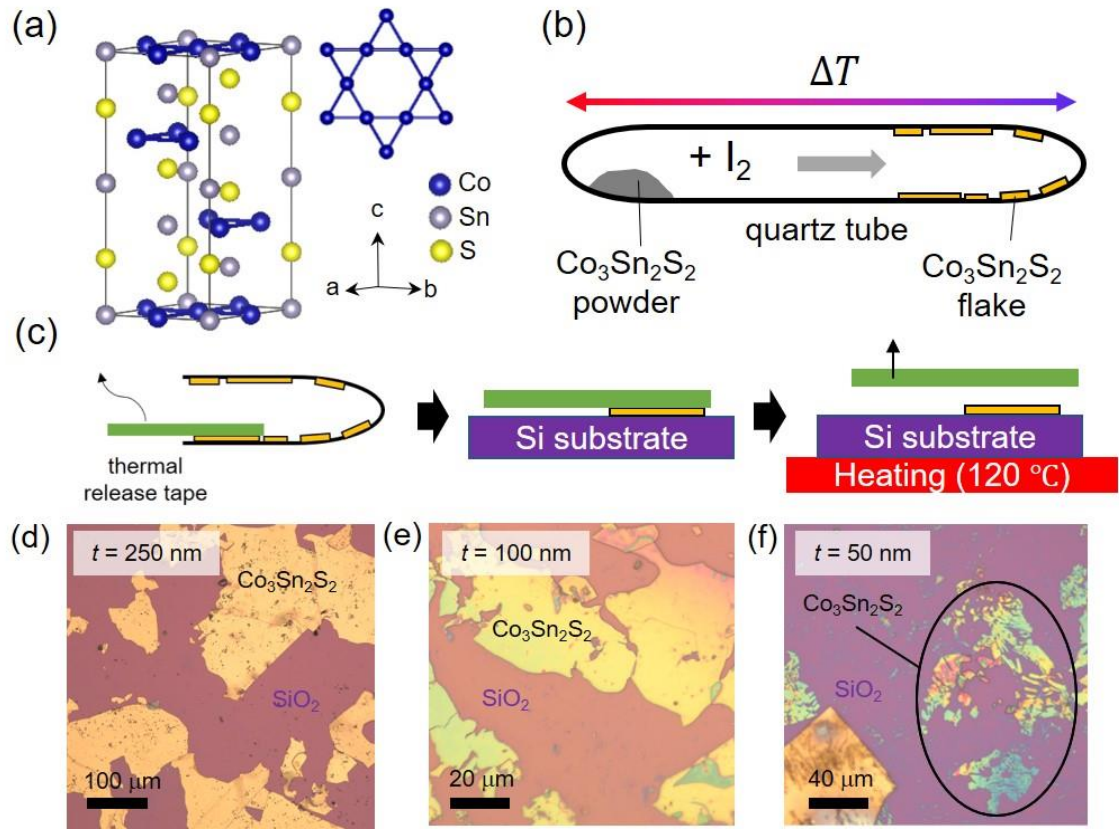

**Figure 1. CVT growth of Co$_3$Sn$_2$S$_2$ thin flakes.** (a) A layered crystal structure of Co$_3$Sn$_2$S$_2$ with 2D Kagome lattice formed by the magnetic Co atoms. (b) A schematic for the set-up of the CVT. (c) Schematics illustrating the transfer process of thin flakes from the quartz tube to the SiO$_2$/Si substrate using thermal release tape. (d)-(f) Optical microscope image of as-grown Co$_3$Sn$_2$S$_2$ thin flakes with a thickness of around 250 nm (d), 100 nm (e), and 50 nm (f). The characteristic violet-blue interference color appears as the thickness becomes small.

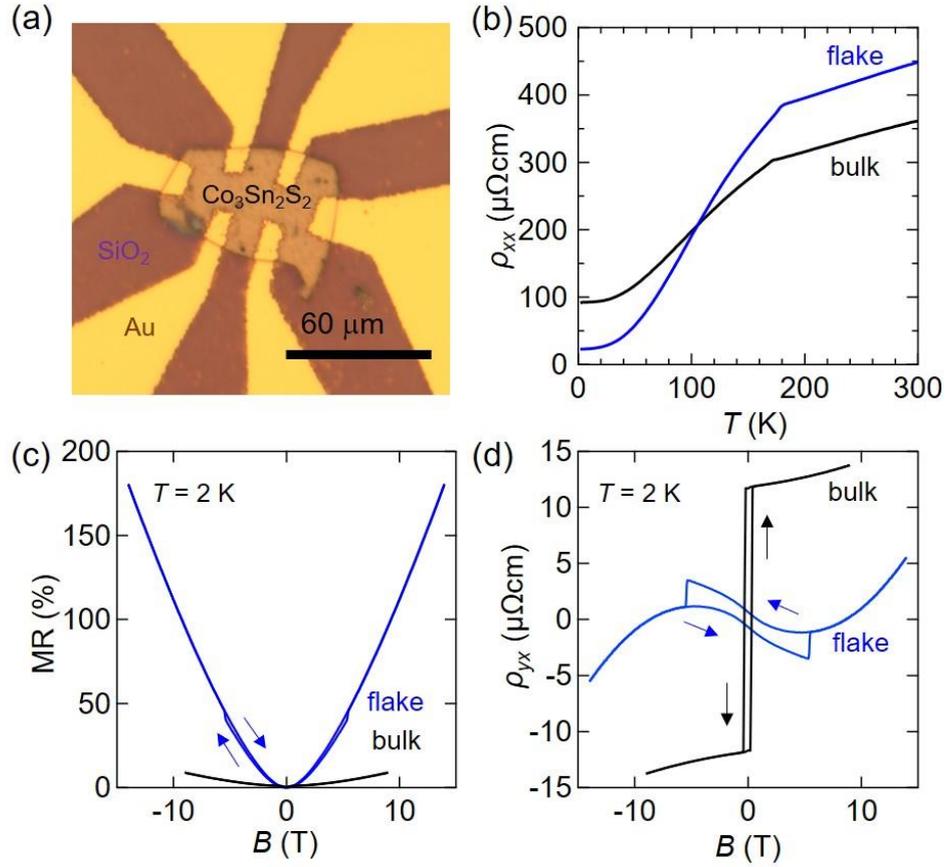

**Figure 2. Resistivity in Co$_3$Sn$_2$S$_2$ thin flake and bulk sample.** (a) Optical microscope image of the thin-flake device with a thickness of 250 nm. (b) Temperature dependence of the longitudinal resistivity at zero magnetic field. (c)(d) Magnetoresistance (MR) (c) and Hall resistivity (d) measured in a transverse magnetic field up to $B$ = 14 T at $T$ = 2 K.

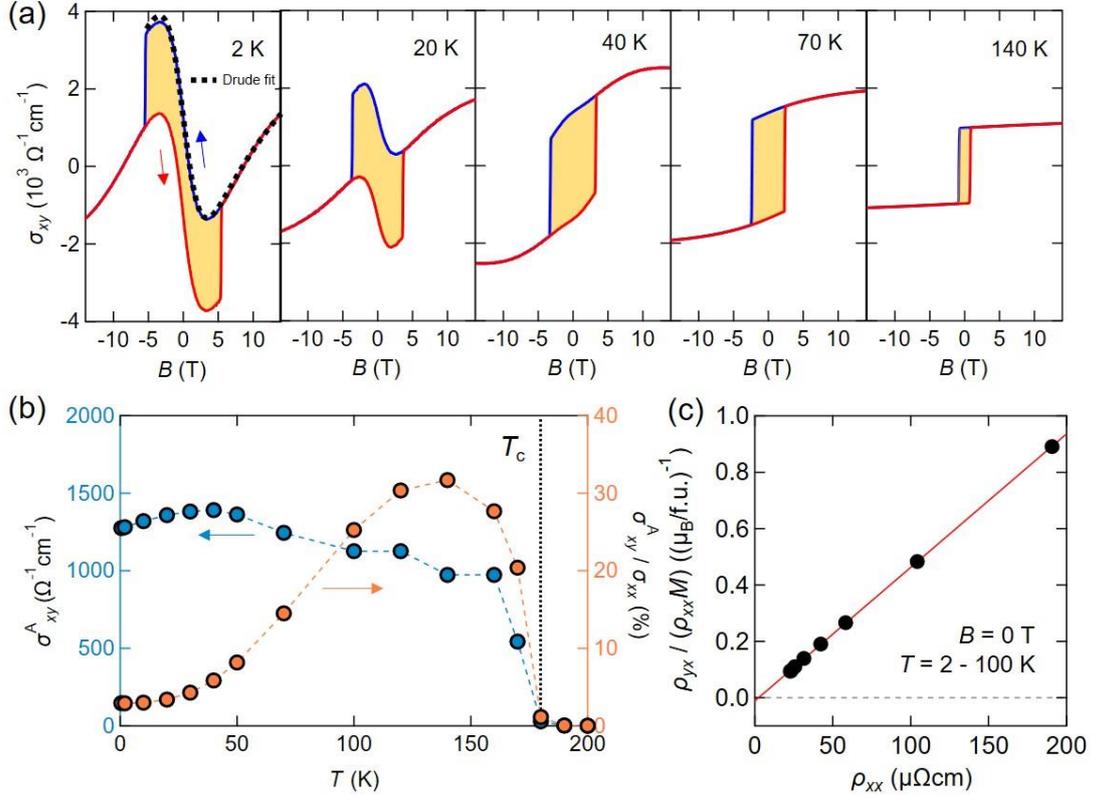

**Figure 3. High electron mobility and large anomalous Hall effect in $Co_3Sn_2S_2$ thin flake.** (a) Magnetic field dependence of the Hall conductivity ($\sigma_{xy}$) at various temperatures, showing coexistence of the AHE with the hysteretic behavior and the ordinary Hall effect (OHE) with the field-nonlinear profile. We attribute the yellow shaded regions to the AHE component while the "dispersive-resonance" profile typical to high mobility systems arises from the OHE. The fitting of the OHE curve with the Drude model is shown by the dashed black line for $T = 2$ K (see main text for detail). (b) Temperature dependence of the anomalous Hall conductivity ($\sigma^A_{xy}$) and the anomalous Hall angle ($\sigma^A_{xy}/\sigma_{xx}$) at zero magnetic field. (c) $\rho_{yx}/(\rho_{xx}M)$ plotted against $\rho_{xx}$ at zero magnetic field for the temperature range of $T = 2$-$100$ K. We used the magnetization ($M$) value of a bulk single crystal (data not shown).

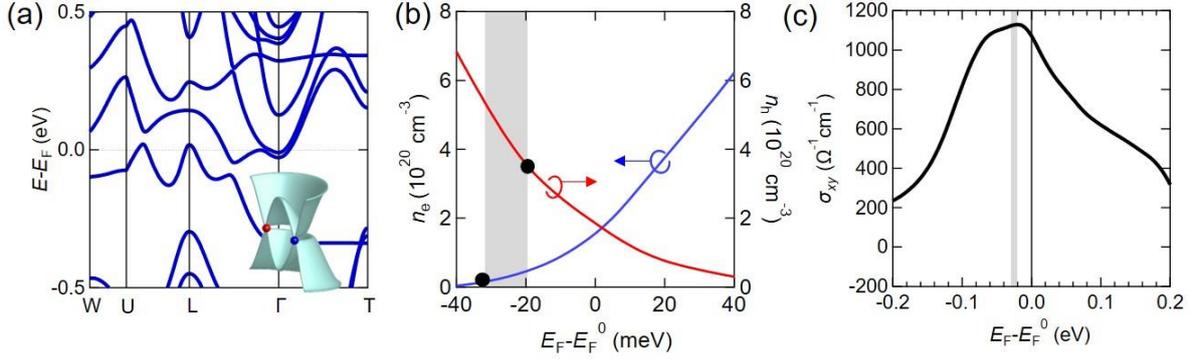

**Figure 4. Band structure calculations of $Co_3Sn_2S_2$.** (a) The band structure of $Co_3Sn_2S_2$ with spin-orbit coupling (SOC) along the high-symmetry paths, which was calculated using the experimental lattice parameters. (b) The calculated carrier density of electron (blue line) and hole (red line) as a function of Fermi energy $E_F$ relative to the original $E_F^0$ value for the pristine bulk compound. The black dots represent the experimental values for electron/hole carrier densities, suggesting the effective shift of $E_F$ to the hole-doping side by 20-30 meV in the thin flake (shaded region). (c) The calculated anomalous Hall conductivity as a function of energy. The maximum appears in a slightly hope-doped (~20 meV) region, which originates from the SOC-gapped nodal ring below the $E_F$. The shaded region represents the Fermi energy of the thin flake estimated from the carrier densities (Fig. 4(b)).

**Table 1.** Summary of the mobility of electron ($\mu_e$) and hole ($\mu_h$), the carrier density of electron ($n_e$) and hole ($n_h$), reported or observed at $T = 2$ K.

| $Co_3Sn_2S_2$ | $\mu_e$ (cm$^2$/Vs) | $\mu_h$ (cm$^2$/Vs) | $n_e$ (/cm$^3$) | $n_h$ (/cm$^3$) | Ref |
|---|---|---|---|---|---|
| bulk (flux) | 730 | 640 | $7.6 \times 10^{19}$ | $9.3 \times 10^{19}$ | [18] |
| bulk (CVT) | 1796 | 1768 | $8.4 \times 10^{19}$ | $8.8 \times 10^{19}$ | [30] |
| bulk (Bridgman) | 574 | 532 | $5.3 \times 10^{19}$ | $4.7 \times 10^{19}$ | [34] |
| thin flake (CVT) | 2587 | 119 | $1.5 \times 10^{19}$ | $5.3 \times 10^{20}$ | This work |